\documentclass[12pt,preprint]{emulateapj}
\usepackage{epsfig}
\usepackage{hyperref}
 
\shortauthors{Garnavich et~al.}
\shorttitle{SN II-P Light Curves with Kepler}
\slugcomment{Accepted for publication in the {\it ApJ}, January 18, 2016}

\newcommand\snIIPa{KSN~2011a}
\newcommand\snIIPd{KSN~2011d}

\begin{document}

\title{Shock Breakout and Early Light Curves of Type II-P Supernovae Observed with Kepler}
\author{P. M. Garnavich\altaffilmark{1}, B. E. Tucker\altaffilmark{2,3}, A. Rest\altaffilmark{4}, E. J. Shaya\altaffilmark{5}, R. P. Olling\altaffilmark{5}, D Kasen\altaffilmark{3,6}, 
A. Villar\altaffilmark{7}}

\altaffiltext{1}{Department of Physics, University of Notre Dame, 225 Nieuwland Science Hall, Notre Dame, IN, 46556-5670, USA.}
\altaffiltext{2} {The Research School of Astronomy and Astrophysics, Australian National University, Mount Stromlo Observatory, via Cotter Road, Weston Creek, ACT 2611, Australia.}
\altaffiltext{3}{Department of Physics and Astronomy, University of California, Berkeley, CA 94720-3411, USA.}
\altaffiltext{4}{Space Telescope Science Institute, 3700 San Martin Drive, Baltimore, MD 21218, USA.}
\altaffiltext{5}{Astronomy Department, University of Maryland, College Park, MD 20742-2421, USA.}
\altaffiltext{6}{Lawrence Berkeley National Laboratory, 1 Cyclotron Road, Berkeley, California 94720, USA. }
\altaffiltext{7}{Harvard-Smithsonian Center for Astrophysics, 60 Garden St., Cambridge, MA 02138, USA.}


\begin{abstract}

We discovered two transient events in the Kepler field with light
curves that strongly suggest they are type~II-P supernovae. Using the
fast cadence of the Kepler observations we precisely estimate the rise time
to maximum for KSN2011a and KSN2011d as 10.5$\pm 0.4$ and 13.3$\pm 0.4$ rest-frame days respectively. Based on fits to idealized analytic models, we find the progenitor radius of KSN2011a (280$\pm 20$~R$_\odot$) to be significantly smaller than that for KSN2011d (490$\pm 20$~R$_\odot$) but both have similar
explosion energies of 2.0$\pm 0.3\times 10^{51}$~erg.

The rising light curve of KSN2011d is an excellent match to that
predicted by simple models of exploding red supergiants (RSG). However, the
early rise of KSN2011a is faster than the models predict possibly due to
the supernova shockwave moving into pre-existing wind or mass-loss from
the RSG. A mass loss rate of $10^{-4}$ M$_\odot$~yr$^{-1}$
from the RSG can explain the fast rise without impacting the optical flux at
maximum light or the shape of the post-maximum light curve.

No shock breakout emission is seen in KSN2011a, but this is likely due to the
circumstellar interaction suspected in the fast rising light curve.
The early light curve of KSN2011d does show
excess emission consistent with model predictions of a shock breakout. This
is the first optical detection of a shock breakout from a type II-P supernova.

\end{abstract}

\subjectheadings{supernova: general --- supernovae: individual (KSN2011a, KSN2011d, SN1999ig) --- shock waves --- stars: mass loss}

\section{Introduction}

Type~II-P supernovae (SNIIP) result from the core-collapse of supergiant stars with significant hydrogen envelopes. Stars exceeding about eight times the mass of the Sun evolve to produce an iron core which can not be supported against gravity and the
resulting collapse drives a shock wave that disrupts the star. Details of how the gravitational energy is converted to an explosion driven by an outward propagating shock is not completely understood, and may require core accretion instabilities
\citep{blondin03} or additional energy deposition by neutrinos \citep{bethe85}. However,
there is clear observational evidence from archival studies of nearby SNIIP that their progenitors are supergiant stars with radii several hundred times that
of the Sun \citep[see][for a review]{smartt15}.

When the shock generated by the core-collapse reaches the surface of the star,
a bright flash of hard radiation is expected \citep{klein78,falk78}. The time-scale for shock breakout is roughly the time it takes light to traverse the stellar radius \citep{nakar10}.
For typical supergiants this timescale is less than an hour, meaning shock breakouts
are very difficult to observe directly. Strong indirect evidence for a hard radiation from a breakout in a SNIIP comes from the ionized circumstellar rings around SN~1987A
\citep{fransson89}. Fortuitously, shock breakouts in two SNIIP have
recently been directly detected using the ultraviolet capabilities of the GALEX satellite \citep{schawinski08, gezari15} . Both of these UV observed shock breakouts lasted significantly longer than an hour implying either the supergiant has an extremely large radius or the presence of circumstellar material prolonged the UV emission \citep{ofek10,chevalier11}.

After shock breakout the bulk of the exploded star expands and the effective temperature drops. The competition between the increasing size of the
photosphere and the falling temperature determines the early light curve on
the time-scale of a few days. For simple assumptions of a fixed density profile
and constant opacity dominated by electron scattering, the photospheric radius and
temperature can be parameterized by the progenitor mass, radius and explosion
energy \citep{nakar10,rabinak11}. Approaching maximum light these simple assumptions break down, and detailed modeling is required to account for opacity variations with
depth and wavelength \citep[e.g.][]{dessart13}. So observations of the early light curve are important in constraining progenitor properties while relying on a minimum of assumptions.

A recent study of the rise time of SNIIP has suggested that their
progenitors are typically smaller than supergiants cataloged in the
Galaxy \citep{gonzalez15}. Some of this discrepancy could be due to
selection bias in the catalogs since larger stars at a given temperature are easier to detect than their smaller cousins. Still, it may be that progenitors of SNIIP are more
compact than thought, or circumstellar interaction makes the rise time
appear shorter than expected.

Here, we present Kepler Space Telescope observations of two SNIIP
candidates. The light curves begin before explosion and
were obtained with unprecedented 30-minute cadence and good photometric
precision. While these Kepler observations have several advantages over
other studies of SNIIP, the red-sensitive Kepler bandpass is not ideal
for detecting shock breakout radiation. Further, the way Kepler data
was taken made it difficult to study transient events in ``real time'', so little is known about these supernovae other than their exquisite light curves which
are analyzed in their entirety by \citet{tucker15}.

\begin{figure}
\epsscale{1.3}
\plotone{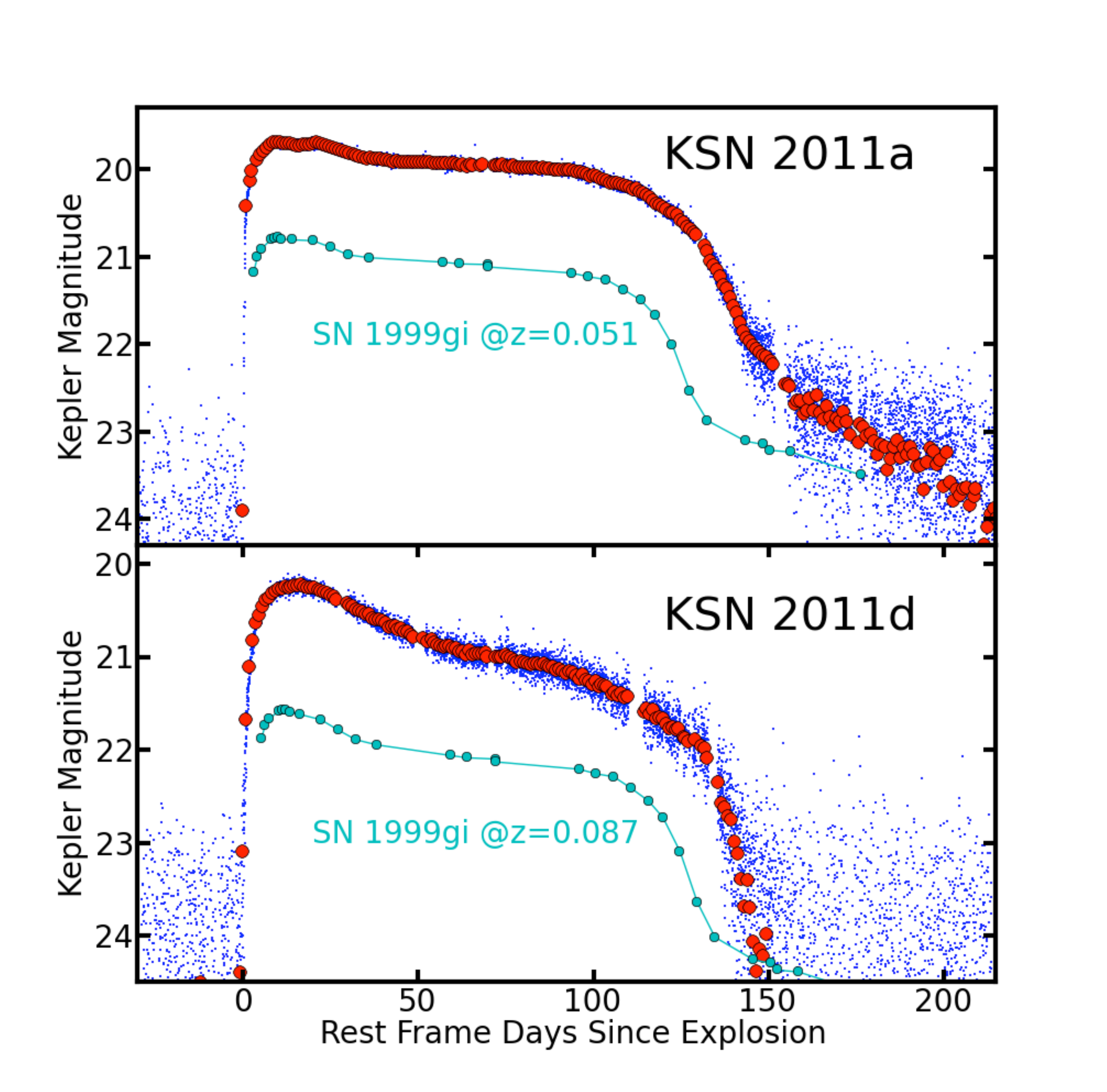}
\caption{The Kepler light curves of KSN~2011a (top) and KSN~2011d (bottom). The
blue points are magnitudes estimated from the standard Kepler 30-minute cadence while the large red symbols show 1-day medians. The small symbols connected by a line
display the light curve of the proto-typical type~IIP SN~1999gi \citep{leonard02} after correction to the redshift of the Kepler events. The initial rise of KSN2011a is clearly faster than KSN2011d based on the number of red points (1-day median) before
maximum light.}
\label{full}
\end{figure}

\section{Observations}\label{sec:obs}

While the primary goal of the Kepler Mission \citep{haas10} was to find and study extra-solar planets, it also provided nearly continuous observations of many galaxies.  Several Kepler guest observer projects monitored about 500 galaxies at 30-min cadence to look for brightness variations in their centers indicative of an active galactic nucleus or to specifically search for supernovae. Targets were selected from the 2MASS extended source catalog (NASA/IPAC IRSA) and the NASA/IPAC Extragalactic Database (NED) . Typically, galaxies were monitored for two to three years leading to the discovery of
three type~Ia supernovae \citep{olling15}, one probable type~IIn event \citep{garnavich16}, and the supernovae presented here. Unfortunately the timescale for release of Kepler data meant that follow-up of the
events was not possible from ground-based observatories. We did obtain spectra of the host galaxies which provide redshifts of the supernovae and information on the
environment around the progenitors \citep{tucker15}.

On a timescale of minutes to hours, Kepler provides photometric precision of a few parts in a million for bright sources. However, on longer timescales, various systematic effects considerably reduce the precision of the standard Kepler products.
For example, the Kepler observations were organized in three-month segments labeled quarters Q0 to Q16. Each quarter the spacecraft rotated to keep the Sun on
the Solar panels resulting in the targets shifting to different detectors. About once per month, the spacecraft goes through a pointing maneuver to downlink the data to Earth. Significant sensitivity variations in the pipeline light curves after re-pointing maneuvers are removed through special processing. Details of our
Kepler reduction procedures can be found in \citet{olling15,shaya15}.

\begin{figure*}
\epsscale{0.8}
\plotone{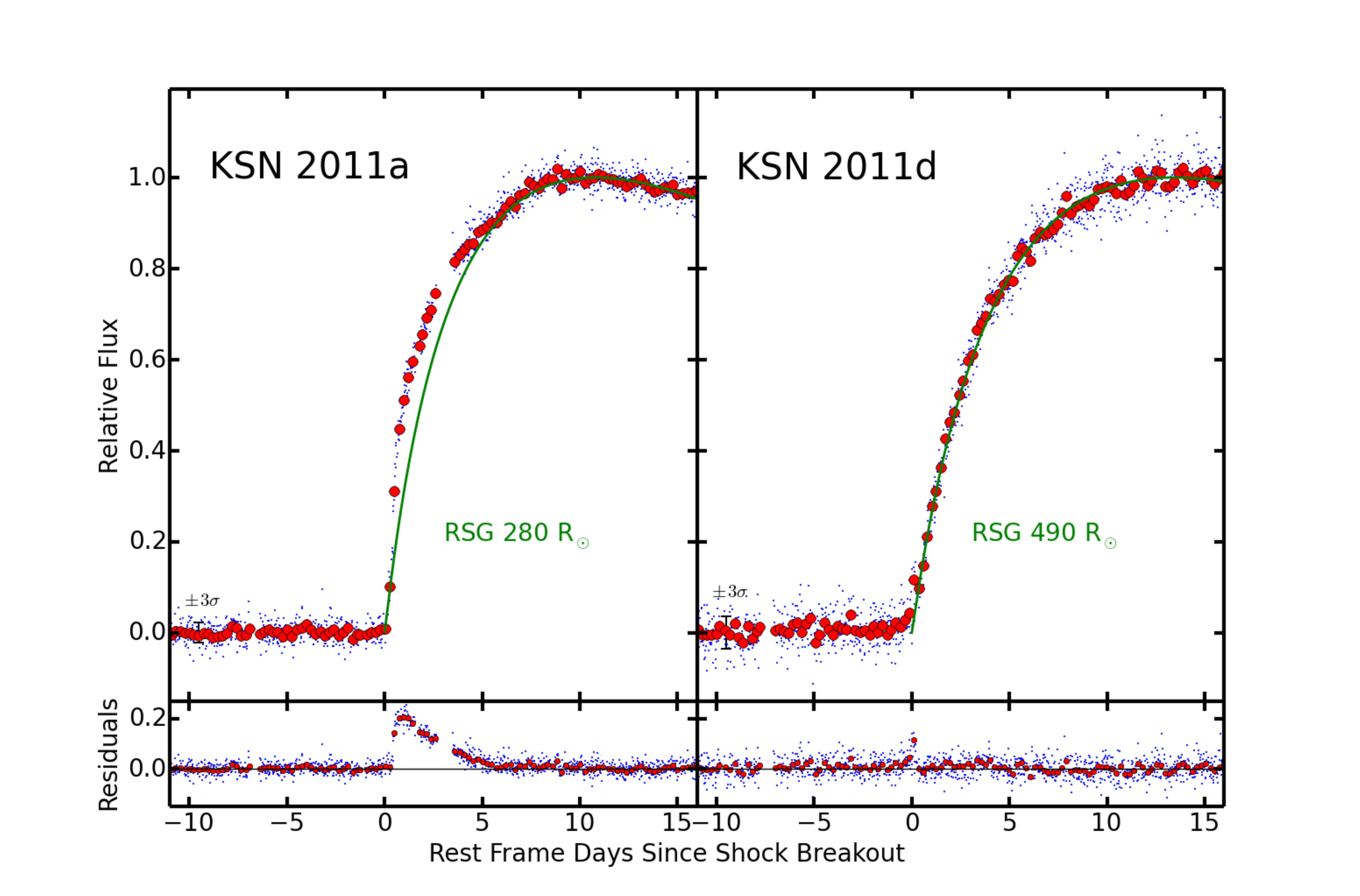}
\caption{The early light curves of the two Kepler type~II-P supernovae.  Blue dots
are individual Kepler flux measurements with a 30~minute cadence and the
red symbols are 6-hour medians. The x-axis shows the redshift corrected time
since shock breakout estimated from the model fit. Note that a shock traversing a red supergiant can take about a day to reach the surface, so we can not measure the
time of core-collapse. {\it Right:} The light curve for KSN2011d with a
model fit assuming a progenitor radius of 490~R$_\odot$. An errorbar at $-10$~days indicates the 3-$\sigma$ uncertainty on the median points. The lower panel shows the
residuals to the fit. {\it Left:} The light curve of KSN2011a with a model fit
using a progenitor radius of 280~R$_\odot$. The model can not match the fast rise
early in the light curve and still fit the time of maximum. The lower panel shows
significant residuals that decay on a timescale of 5 days.}
\label{rise}
\end{figure*}

\section{Light Curves}\label{sec:lightcurves}


\snIIPa\ was discovered in the galaxy KIC8480662 which is a bright 2MASS galaxy
at a redshift of $z=0.051$ \citep{tucker15}. The Kepler light curve shows a fast rise, a broad
maximum followed by a long plateau (see Fig.~\ref{full}). Finally there is
a rapid decay followed by an exponential decline. The light curve is characteristic
of SNIIP.


\snIIPd\ was discovered in the galaxy KIC10649106 which is also a 2MASS
cataloged galaxy at a redshift of $z=0.087$ \citep{tucker15}. Its light curve
also shows a fast rise, a broad maximum and then a slow decay before falling
off the ``plateau'' after 130 days. KSN2011d appears to fade faster
on the plateau than KSN2011a, but part of that can be attributed to the higher
redshift which means the bandpass contains bluer light that fades more
quickly in SNIIP. A detailed analysis of the full light curves
can be found in \citet{tucker15}.

These Kepler supernovae light curves are very similar to several well-observed SNIIP events such as SN1999em (Suntzeff, private com.), SN1999gi \citep{leonard02} and SN2012aw \citep{bose13}. The Kepler supernovae are at significantly higher redshifts than
these local events, so k-corrections are important, but there
is no color information for the Kepler events. Therefore, we
use the $BVRI$ magnitudes of the nearby supernovae to correct them
to the Kepler observed frame.

For the nearby supernovae we create a spectral energy distribution (SED)
for each epoch observed in multiple filters. Missing filters are
interpolated from adjacent epochs. The SEDs are corrected for Milky Way extinction using \citet{schlafly11}. The SED is
corrected to the redshift of the Kepler events, convolved with the Kepler bandpass and the result is integrated to give the total photon flux. The result is also reddened to match the Milky Way extinction in the direction of the Kepler supernova. Kepler magnitudes are in the AB system, so the Kepler bandpass is convolved
with a spectrum with constant $F_\nu = 3631$~Jy and integrated
to determine the magnitude zeropoint.

For comparison, Figure~\ref{full} displays the light curve of the well-observed local event SN~1999gi after correction to the redshifts of the Kepler supernovae. SN~1999gi was a slightly
fainter than typical SNIIP \citep{bose13}, and ignoring unknown host extinction, it is over a magnitude fainter than these two Kepler supernovae. Still, the shape of the
light curve and length on the plateau make the SNIIP classification of KSN2011a and KSN2011d very solid.

\section{Analysis}\label{sec:analysis}

\subsection{Rise to Maximum}

The rapid cadence of the Kepler observations provide a unique window on the
early rise of supernovae. In particular, SNIIP have rise times
on the order of a week and are difficult to capture in typical ground-based
surveys \citep[e.g.][]{gonzalez15}. In Figure~\ref{rise}, we show the Kepler
light curves beginning several days before explosion and ending soon
after maximum light. There are approximately 500 individual photometric
measurements between the initial brightening and maximum light. But these
events are at significant distance, so we have combined the Kepler cadence
into 6-hour median bins to improve the signal-to-noise ratio of the light curves.

\begin{figure*}
\epsscale{1.1}
\plottwo{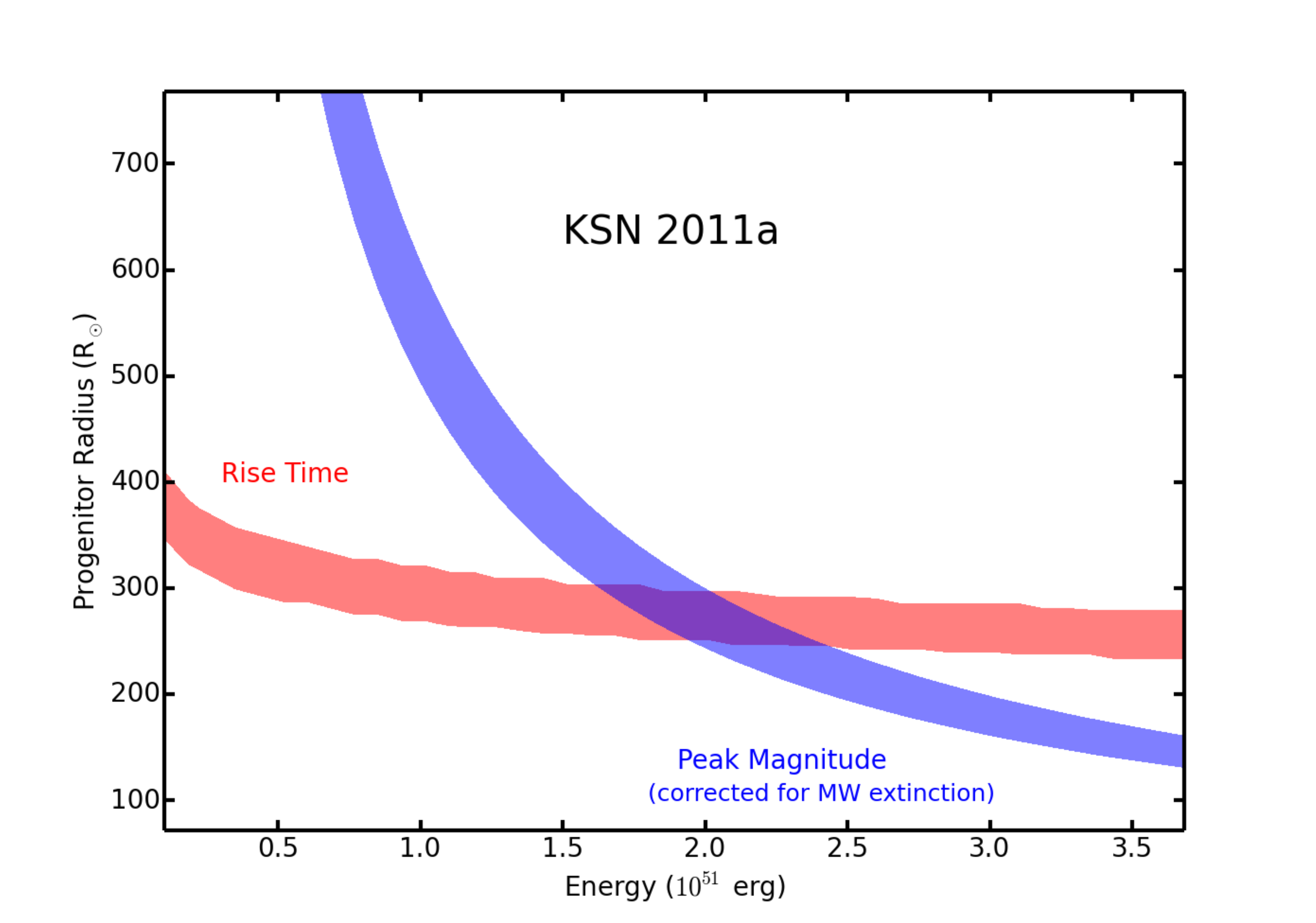}{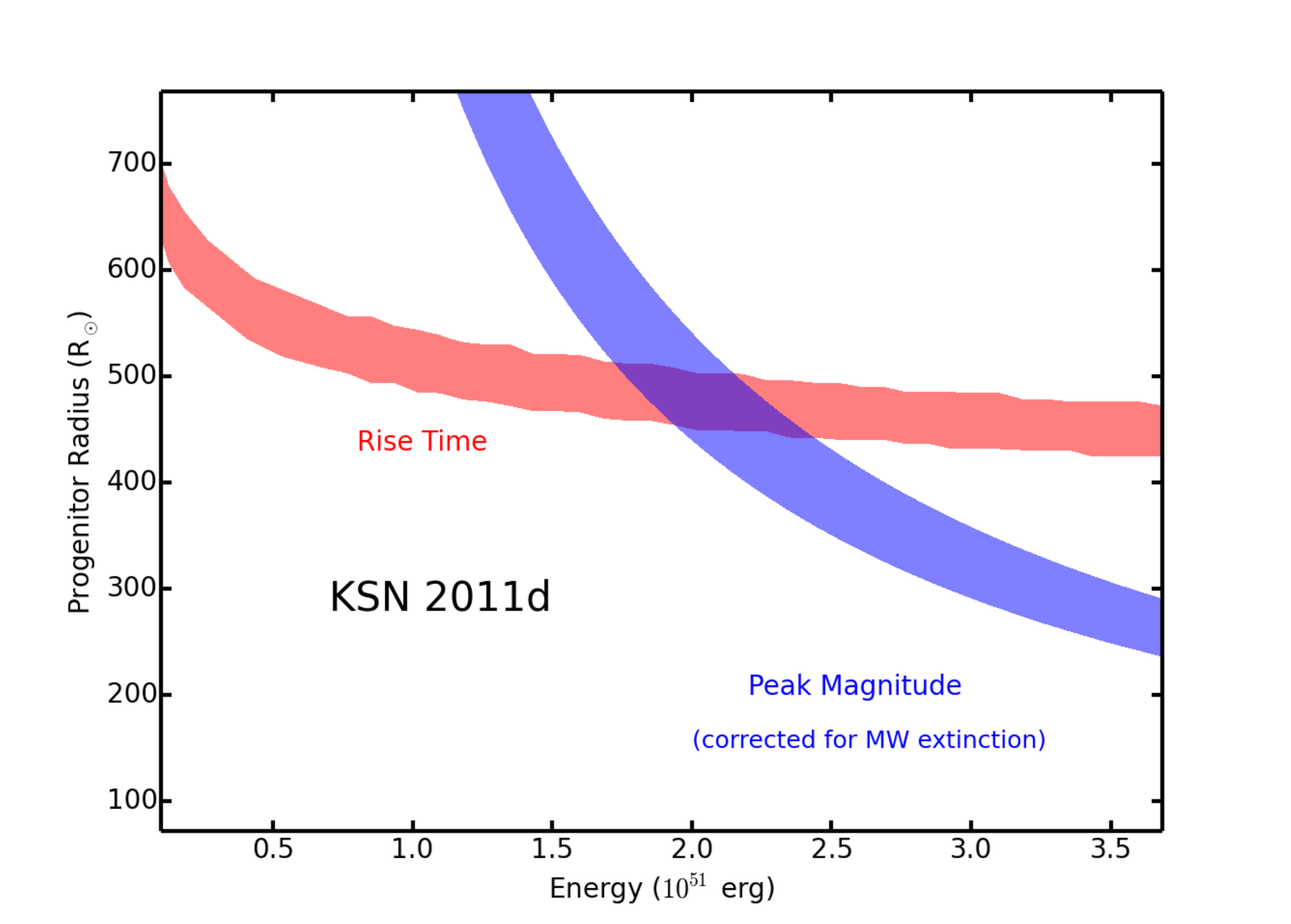}
\caption{Constraints on the progenitor size and explosion energy based on the observed
light curve rise time and peak magnitude. In the \citet{rabinak11} models for red supergiants, the rise time of the light curve (red) is good at constraining the progenitor size, while the peak brightness (blue) is best at limiting the range of explosion energies. The widths of
the bands represent the uncertainties in the observed quantities. The supernova luminosities have been corrected
for Milky Way extinction but no host reddening has been assumed.
Here a progenitor mass of 15~M$_\odot$\ is assumed, but the results are not sensitive to mass for the range of 10 to 20 M$_\odot$. {\it Left:} For KSN2011a the intersection of the rise time and peak brightness suggests a progenitor radius of 280$\pm 20$
R$_\odot$ and an explosion energy of 2.0$\pm 0.3$~B. {\it Right:} The Kepler
observations of KSN2011d suggest a larger progenitor at 490$\pm 20$ R$_\odot$
but a similar explosion energy of 2.0$\pm 0.3$~B.}
\label{models}
\end{figure*}

To fit the pre-maximum rise of the Kepler events, 
we calculated a grid of light curves using
the \citet{rabinak11} red supergiant (RSG) model.
The RSG model assumes a power-law density structure with
an index $n=3/2$.
We vary the progenitor radius and explosion
energy keeping the  stellar mass
at 15~M$_\odot$ which is typical for core-collapse supernovae 
\citep{smartt15}. We assume fully ionized hydrogen envelopes
($\kappa=0.34$ cm$^2$~g$^{-1}$) and the set the normalization
of the ejecta density to $f_p=0.1$, although the results are not sensitive to
this parameter. 
The \citet{rabinak11}
model allows us to calculate the temperature
and radius of the expanding photosphere as
a function of time. For each epoch we
construct a blackbody corrected to the
redshift and distance of the Kepler supernova
and multiply the spectrum by the Kepler
sensitivity function. We then integrate
and normalize the resulting flux using
the zeropoint calculated in the AB magnitude
system. For each pair of initial radius and
explosion energy we have a light curve
from which we derive a rise time
and peak magnitude.

The observed rise time (and uncertainty)
defines a band in the radius
versus energy plane that is nearly
horizontal while the observed magnitude (and
error) defines a band that cuts diagonally
across the parameters of interest. The results for the two
Kepler supernovae are presented in Fig.~\ref{models}.
The intersection of the two bands is a consistent
fit to the rise time and peak brightness (corrected for
Milky Way extinction)  
and tightly constrains the derived quantities of
progenitor radius and explosion energy.
We also created model grids for
progenitor masses of 10 and 20 M$_\odot$,
but the resulting parameters did not
differ significantly from the 15~M$_\odot$
calculations.

The model light curves that best match
the observed rise time and peak magnitude
are then fit to the actual light curves
by minimizing the residuals between
the model and data. From this a new
time of explosion is estimated
and an improved rise time is calculated.
This iterative method converges quickly
and the resulting fits are shown in
Fig.~\ref{rise}.

For KSN2011d, the \citet{rabinak11} model
with a progenitor radius of 490~R$_\odot$
and explosion energy of 2$\pm 0.3$~B\footnote{1 B = 1 foe = $10^{51}$ erg} not only
matches the rise time and peak magnitude,
but the \citet{rabinak11} model predicts the
overall shape of the rise very well. The $\chi^2$
parameter of the best fit model is 1532 for 
1142 free parameters or a reduced $\chi^2_\nu = 1.34$.

The best fit rise time for KSN2011a is 280$\pm 20$~R$_\odot$
and the explosion energy is also 2$\pm 0.3$~B. \citet{kasen09} found
that only a few nearby SNIIP events had energies larger than 1.5~B,
so finding two Kepler supernovae with explosion energies of 2~B
is surprising. But the Kepler supernova search is likely biased toward
discovering luminous events near the limiting magnitude of the survey and
these will tend to have higher explosion energies than supernovae found in volume
limited searches of nearby events.

No correction has been made for possible host extinction because colors are
not measured for these events. The explosion energy will be the parameter most affected
by our uncertainty in host reddening and in our analysis we are
actually estimating its lower limit. \citet{poznan09} estimated the color excess for
forty SNIIP events and from that sample we infer a median visual extinction of
0.80~mag. We note that this sample of relatively nearby SNIIP may not be
representative of the magnitude-limited Kepler discoveries, but it does suggest
significant extinction is not unusual for SNIIP.

The shape of the KSN2011a light curve is
not as well fit by the \citet{rabinak11} 
prediction (left panel of Fig.~\ref{rise}) even
when the time to maximum and
peak magnitude are well matched. In the first
five days the
KSN2011a light curve rises significantly faster
than the model even though the same physics
resulted in an
excellent fit to the KSN2011d light curve.
Assuming a smaller progenitor radius does make
the model rise faster, but that model will
then peak much earlier than the observed
10.5 days. Smaller assumed radii also produce
a poor fit near maximum light for blue supergiant
(BSG) models which differ from RSG in their density
profile.

The rapid rise in KSN2011a and the ``extra light''
above the \citet{rabinak11} model photosphere
suggests the supernova shock continued
to propagate into circumstellar material allowing
it to convert
more kinetic energy into luminosity and diffuse
the shock breakout over a longer time.
Strong circumstellar
interaction has been successful in explaining very
luminous events \citep{ofek10,chevalier11}, but
progenitors with low mass loss rates may also
see their early light curves enhanced with a weak
shock interaction \citep{moriya11}.

\citet{moriya11} calculations show that mass loss
rates of order $10^{-4}$ M$_\odot$~yr$^{-1}$ will
cause the early light curve to rise
faster than a bare RSG, while not strongly
affecting the optical peak luminosity or
the light curve during the plateau. Mass-loss rates less than
$10^{-4}$ M$_\odot$~yr$^{-1}$
mean that the circumstellar medium near the
progenitor radius is too low density to
become optically thick when the shock hits it,
so the presence of the wind would have no significant impact on
the light curve even during the early rise. We therefore expect that the mild
interaction seen in KSN2011a is due to
a mass-loss rate just above the $10^{-4}$ M$_\odot$~yr$^{-1}$ threshold.

\begin{figure*}
\epsscale{0.8}
\plotone{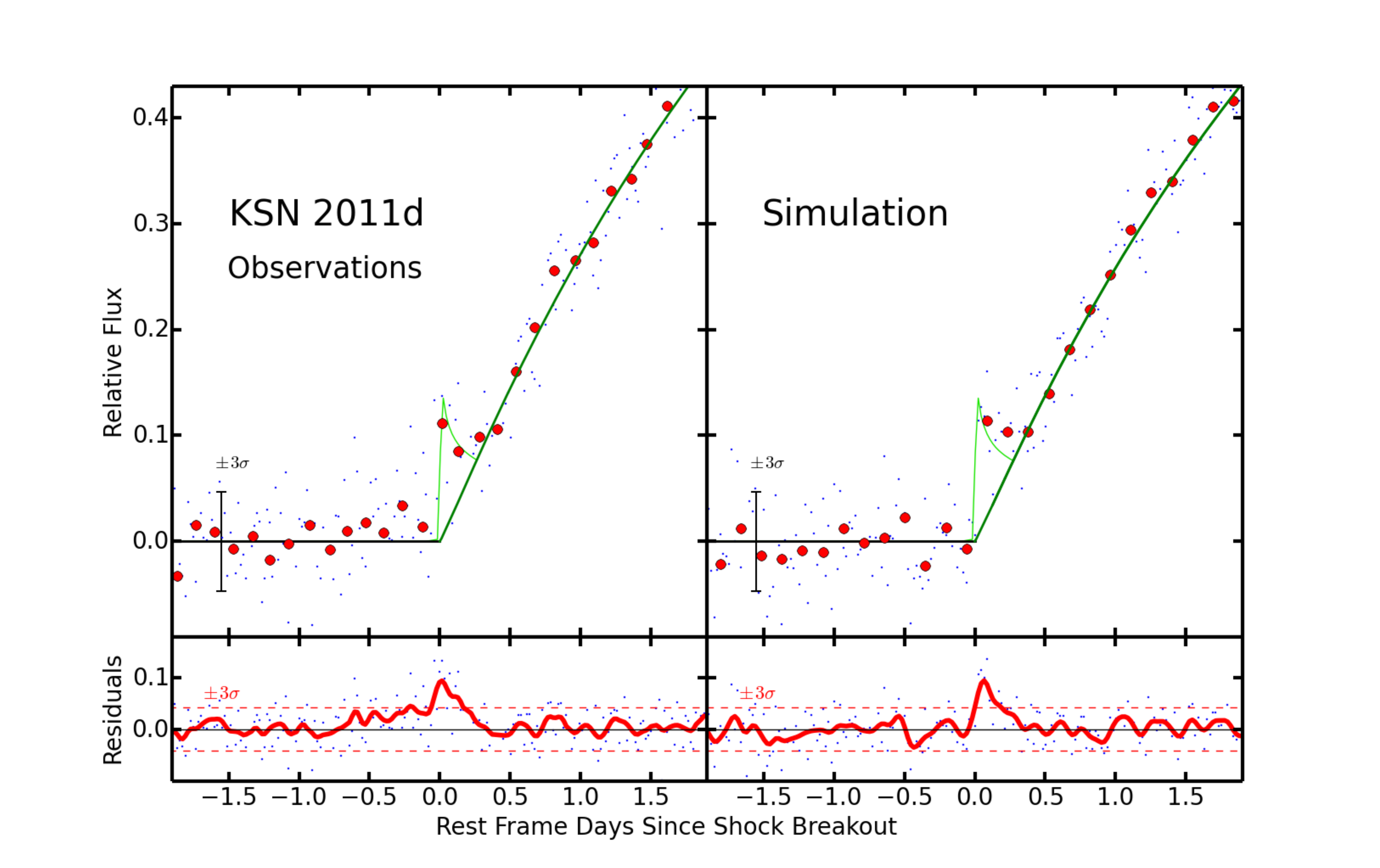}
\caption{{\it Left:} The Kepler light curve of KSN2011d focused on the time expected for shock breakout. The blue dots are individual Kepler measurements and the red symbols show 3.5-hour medians of the Kepler data. An errorbar at $-1.5$~days indicates the 3-$\sigma$ uncertainty on the median points.The green line shows the best fit photospheric model light curve. The lower panel displays the residuals between the observations and the model fit. The thick red line is a Gaussian smoothed residual light curve using a full-width at half-maxmimum of two hours. The dashed red lines indicate 3$\sigma$ deviations of
the Gaussian smoothed curve. The residual at the time expected for shock breakout is more than 5$\sigma$ implying that the feature is unlikely to be a random fluctuation. {\it Right:} A simulated light curve created using the statistical properties of the Kepler photometry and the best fit photospheric model. In addition,
a \citet{nakar10} shock breakout model (light green line) for an explosion energy of 2~B and radius of 490~R$_\odot$ is compared with both the data and simulation.}
\label{zoom}
\end{figure*}

\subsection{Shock Breakout}

\subsubsection{KSN2011a}

The fast cadence and continuous coverage of Kepler should,
in principle, allow us to see the shock breakouts in
these SNIIP. In the case of KSN2011a where we suspect 
circumstellar interaction, the shock breakout was 
likely reprocessed over the diffusion time in the optically thick
wind. The shock continued into the wind and converted additional kinetic energy
into luminosity that we see as excess flux during the photospheric rise. 
The peak absolute Kepler magnitude from the circumstellar interaction is $M_{Kp}=-15.5$
mag, but much of the total energy is likely emitted at shorter wavelengths.

\subsubsection{KSN2011d}

In the KSN2011d light curve (Fig.~\ref{rise}), there is 
a single 6-hour median flux point that deviates from
the light curve model by 4 standard deviations ($\sigma$) at the time expected
for shock breakout. A close-up of this time period 
is shown in Fig.~\ref{zoom} using a binning width of 3.5 hours. Extrapolating the
\citet{rabinak11} photospheric model to zero flux predicts shock breakout at
$t_0 = 2455873.75\pm 0.05$~BJD which corresponds to the time of the largest deviation
from the model.

\begin{deluxetable*}{lccccccc}
\tabletypesize{\scriptsize}
\tablecaption{Kepler Type~II-P Supernova Candidates \label{table1}}
\tablewidth{0pt}
\tablehead{
\colhead{Name$^a$} & \colhead{Host} & \colhead{SN} & \colhead{Redshift} & \colhead{MW A$_V$} & \colhead{Peak Kp$^c$} & \colhead{Date of Breakout} & \colhead{Rise Time} \\ 
\colhead{} & \colhead{KIC$^b$} & \colhead{Type}  & \colhead{($z$)} & \colhead{(mag)}
& \colhead{(mag)} & \colhead{(BJD-2454833.0)} & \colhead{(days)}
}
\startdata
KSN 2011a & 08480662 & IIP & 0.051 & 0.194 & 19.66$\pm 0.03$ & 934.15$\pm 0.05$ & 10.5$\pm 0.4$ \\
KSN 2011d & 10649106 & IIP & 0.087 & 0.243 & 20.23$\pm 0.04$ &  1040.75$\pm 0.05$ & 13.3$\pm 0.4$
\enddata
\tablenotetext{a}{Kepler SuperNovae (KSN) 2011b, 2011c and 2012a were published in
\citet{olling15}}
\tablenotetext{b}{Kepler Input Catalog \citep{brown11}}
\tablenotetext{c}{not corrected for extinction}
\tablenotetext{ }{ }
\tablenotetext{ }{ }
\end{deluxetable*}

When we subtract the best fit photosphere model for KSN2011d there remains
seven Kepler photometric observations within five hours of $t_0$
that are 3$\sigma$ or more above
zero (lower panel in Fig.~\ref{zoom}). To avoid bias that might come
from dividing the data into bins, we have smoothed the light curve residuals using
a Gaussian with a full-width at half-maximum (FWHM) of 3 hours. There is a clear
6$\sigma$ peak at the time expected for breakout and we
conclude that this is, indeed, the shock breakout from KSN2011d. The
shock breakout flux is 12\%\ of the peak flux of the supernova, corresponding
to a Kepler magnitude of 22.3$\pm 0.2$ after correcting for Milky Way extinction.

In the \citet{nakar10} shock breakout model, the initial rise is the result of diffusion of the shock emission before the shock reaches the stellar surface
and is only of order five minutes. This is too short a time for even the Kepler
cadence, so the rise to shock breakout is unresolved. After shock breakout the flux decay follows a $t^{-4/3}$ power-law in time until the expanding photosphere
dominates the luminosity. This decay is relatively slow and allows the
breakout to remain detectable for several hours. From the \citet{nakar10}
formulation, we can estimate the ratio between the peak flux from the
shock breakout, $F_{SB}$, and the maximum photospheric flux, $F_{max}$,
which we approximate as the brightness 10~days after explosion. Using the ratio between the shock peak and photosphere maximum is particularly useful since it eliminates the uncertainty caused by distance and dust extinction. In the rest-frame optical (5500\AA) the flux ratio is
\begin{equation}
F_{SB}/F_{max}\: =\; 0.25\; M_{15}^{0.54}\; R_{500}^{0.73}\; E_{51}^{-0.64} \label{eq1}
\end{equation}
where $M_{15}$ is the progenitor mass in units of 15~M$_\odot$, $R_{500}$ is
the progenitor radius in units of 500~$R_\odot$, and $E_{51}$ is the explosion
energy in units of $10^{51}$~erg. So we expect the shock breakout in a typical RSG
to peak at about 25\%\ of visual brightness of the supernova at maximum. 

Applying the \citet{nakar10} model to KSN2011d (radius of 490~R$_\odot$,
energy of 2~B and a progenitor mass of 15~M$_\odot$), predicts
a breakout temperature of 2$\times 10^5$~$^\circ$K, and equation~\ref{eq1}
gives $F_{SB}/F_{max} = 0.16$, meaning the shock should be
2~mag fainter in the optical than the supernova at maximum. The Kepler
30~minute cadence will smooth the sharp peak of the breakout and lower
the maximum by 20\%\, so
we expect the ratio to be $F_{SB}/F_{max} = 0.13$.
The excess flux seen in Fig.~\ref{zoom} peaks at a relative flux of 0.12$\pm 0.2$
and is consistent with the \citet{nakar10} prediction. 

We use a blackbody spectrum to extrapolate the shock breakout
flux down to the optical and this is likely a poor approximation. So it is surprising that
the semi-analytic model of \citet{nakar10} works so well in matching the
observed breakout. \citet{tominaga11} calculated realistic spectra at breakout
for a variety of RSG models. While color temperatures and integrated luminosities
varied greatly, the peak optical flux at breakout was fairly consistent: between $2\times 10^{37}$ to 1$\times 10^{38}$ erg~s$^{-1}$~\AA$^{-1}$, corresponding to absolute magnitudes between $-14.2$ and $-15.9$ mag. The \citet{tominaga11} model for a 15~M$_\odot$, 1~B and 500~R$_\odot$ RSG predicts a peak at M$_{Kp}=-14.4$
\footnote{Kp$\approx 0.2 g+0.8 r$ where $g$ and $r$ are SDSS
magnitudes (Kepler Calibration webpage)} mag. Doubling the explosion energy would brighten the breakout by about 0.2~mag, yielding a luminosity of M$_{Kp}=-14.6$ mag. The observed shock breakout from KSN2011d is
M$_{Kp}=-15.6\pm 0.3$ mag (after correction for Milky Way extinction; assuming H$_0$=70
km~s$^{-1}$~Mpc$^{-1}$).  Overall, the models
do an excellent job in predicting the optical brightness of the shock breakout
in KSN2011d.

We have simulated the light curve of KSN2011d using the statistical
properties of the Kepler photometry based on 100 days prior to the
supernova detection (right panel in Fig.~\ref{zoom}). The photospheric
rise of the light curve uses the \citet{rabinak11} model and the shock breakout
uses the \citet{nakar10} model with our best fit parameters. Assuming Gaussian
statistics, we created a Monte Carlo simulation of the models sampled at the
Kepler cadence.  The simulation matches the observed light curve extremely well and we again
conclude that we have detected the shock breakout in KSN2011d.

\subsubsection{Radiative Precursor?}

In the 12 hours {\it before} shock breakout there are only two of
the 24 Kepler observations that fall below the median pre-supernova flux. That is,
the light curve of KSN2011d shows a possible slow rise in brightness
starting 0.5~days before breakout. This is intriguing as core-collapse likely occurred nearly a day before shock breakout \citep{chevalier11}. The shock
travels more slowly than photons diffuse through the RSG envelope, allowing
evidence for the shock to reach the surface before breakout.
However, \citet{nakar10}
predicts the shock energy would leak out through diffusion on
a time-scale of only five minutes for a RSG. In contrast \citet{schawinski08}
suggests a ``radiative precursor'' due to photon diffusion could begin
hours before shock breakout and there is some evidence for a precursor seen in the
GALEX detection of a SNIIP five hours before the peak breakout emission. 

While it is tantalizing to claim precursor emission in KSN2011d, the
smoothed flux remains at 3$\sigma$ or less from median value, so this
detection is not definitive. More observations of SNIIP with extremely fast cadence
are needed to determine the diffusion time-scale before shock breakout.

\section{Conclusion}\label{sec:results}

We discovered two transient events in the Kepler field with light
curves that strongly suggest they are SNIIP events. From the
fast cadence of the Kepler observations we determine the time the
supernova shock reached the surface of the progenitor with a precision
of better than 0.1~days. We find the rise time to maximum was 10.5$\pm 0.4$
rest-frame days for KSN2011a and 13.3$\pm 0.4$ days for KSN2011d. From the
rise times
combined with their peak luminosities (not corrected for host extinction),
we estimate the progenitor radius of KSN2011a (280~R$_\odot$) to be significantly smaller than that for KSN2011d (490~R$_\odot$) but both have similar
explosion energies of 2~B.

Our directly measured rise-times for both Kepler events are many
standard deviations larger than the median rise-time of SNIIP estimated from
an ensemble of light curves by \citet{gonzalez15}. This difference
results from the variation of rise-time with effective wavelength. 
\citet{gonzalez15} referenced their
rise-time to the SDSS-$g$ band while light curves peak several days
later in the redder Kepler bandpass. The median radius of
the progenitors estimated by \citet{gonzalez15} is 320~R$_\odot$
which is comparable to the radii we infer from the Kepler data.
As with the \citet{gonzalez15} estimates, radii measured from SNIIP
light curves tend to be at the
low end of the distribution of radii estimated for RSG directly observed
in the Milky Way \citep{levesque05}. Some of this discrepancy may be due
to the idealized analytic models failing to account for opacity changes near
maximum light \citep[e.g.][]{rubin15}.

The rising light curve of KSN2011d is an excellent match to that
predicted by the \citet{rabinak11} models for RSG. However, the
\citet{rabinak11} models underestimate the brightness of KSN2011a during
the first five days and we suggest that the additional flux is due to
the supernova shock moving into a pre-existing wind or mass-loss from the RSG. \citet{moriya11}
has shown that a mass loss
rate of $10^{-4}$ M$_\odot$~yr$^{-1}$ will steepen the light curve
while not strongly impacting the flux at maximum light or the
shape of the post-maximum light curve.

No fast shock breakout emission is seen in KSN2011a, but this is likely due to the
circumstellar interaction suspected in the early light curve.
KSN2011d does show
excess emission at the time expected for shock breakout with a brightness
of 12\%\ that of supernova peak in the Kepler band. The time-scale and brightness
observed for the breakout is consistent with model predictions.

The diversity in the rising light curves of SNIIP observed by Kepler show that early observations are critical in understanding the progenitors and circumstellar
environments of exploding RSG stars.

\acknowledgements

This research has made use of the NASA/IPAC Extragalactic Database (NED) which is operated by the Jet Propulsion Laboratory, California Institute of Technology, under contract with the National Aeronautics and Space Administration. This research has made use of the NASA/ IPAC Infrared Science Archive, which is operated by the Jet Propulsion Laboratory, California Institute of Technology, under contract with the National Aeronautics and Space Administration. This work was partly supported by Kepler grants NNX12AC89G and NNX11AG95G. DK is supported by a DOE office of nuclear physics early career award.

\end{document}